# Lubrication effects on droplet manipulation by electrowetting-on-dielectric (EWOD)

Ken Yamamoto[1, 2, (a], Shimpei Takagi[3, (a], Yoshiyasu Ichikawa[2, 3], Masahiro Motosuke[2, 3]

[1] Department of Earth and Space Science, Graduate School of Science, Osaka University
[2] Water Frontier Research Center (WaTUS), Tokyo University of Science
[3] Department of Mechanical Engineering, Tokyo University of Science
[(a] These authors contributed equally.
e-mail: yam@ess.sci.osaka-u.ac.jp

**Abstract**
Electrowetting has a potential to realize stand-alone point-of-care (POC) devices. Here we report droplet-migration characteristics on oil-infused electrowetting-on-dielectric (EWOD) substrates. We prepare sparse micropillars to retain the oil layer in order to exploit the layer as a lubricating film. A physical model of the droplet velocity is developed, and effects of the lubrication, the oil viscosity, the droplet volume, and the thickness of solid and liquid dielectric layers are discussed. It is found that the droplet velocity is scaled as $U \approx E^2$, which differs from a relationship of $U \approx E^3$ for droplets sliding down on liquid-infused surfaces by gravity. Furthermore, our device achieves droplet velocity of 1 mm s$^{-1}$ at the applied voltage of 15 V. The velocity is approximately tenfold as high as the same condition (applied voltage and oil viscosity) on porous-structure-based liquid-infused surfaces. The achieved high velocity is explained by a lubrication-flow effect.

## I. INTRODUCTION

The wettability of a solid surface can be changed by applying voltage. This phenomenon is called the electrowetting (EW). It is exploited for some tunable lenses, prisms, and optical switches[1–6], and its application was expanded in the 2000s to biochemical applications called digital microfluidics[1,7–11], as electrowetting-on-dielectric (EWOD) was developed. This is because EWOD, which inserts a dielectric layer in-between droplets and electrodes, can manipulate droplets outside mirofluidic channels and in real time[12]. Moreover, EWOD is a robust system because it does not contain moving parts. These features, as well as its high connectivity to other technologies used in biochemical processes, opened the possibility of the droplet-based biochemical analysis on demand.

As our society has progressed to specialize medical care and tests for individual (point-of-care, POC)[11], the programmable nature of EWOD droplet manipulation has many advantages. POC biosensors are also expected to be a powerful tool under pandemic situations because a possible infected individual can test by oneself while avoiding a contact with others. On the other hand, there are many requirements (e.g., simplicity, cost, rapidity, portability, and sensitivity) for the POC devices, and they often conflict each other. For instance, nucleic acid tests are highly sensitive but need multiple steps; thereby realizing it with paper-based (simple and cost-effective) devices is challenging[13]. From this viewpoint, EWOD devices are suitable for automatically processing multiple steps, although they have problems to be solved, such as fabrication cost[14], reliability[15], portability[11], and high-voltage requirement (typically $10^2$ V)[7,11,16]. Especially, driving voltage is one of the crucial points because access to a power source is frequently limited for the POC devices[13,17].

There are several ways to lower the driving voltage of EWOD: using highly-dielectricmaterials[18], decreasing the resistance[6,19–21], using nanofluids[22], using self-healing $Al_2O_3$ dielectrics[23], and so on. Moon et al.[18] employed a very thin (700 Å) and high-dielectric-constant (~180) layer and confirmed the transport of a droplet (460 nL) with 15 V. Yi et al.[19] achieved to lower the manipulation voltage to 65 V with a thin dielectric layer coupled with a hydrophobic coating. Liquid-infused surfaces (LISs) that achieves negligible contact angle hysteresis by infusing lubricating fluid into micro- and/or nano-textured surfaces[24–28] are also used to decrease the applied voltage further. Hao et al.[6] used oil-infused PTFE membrane and achieved low-voltage actuation by eliminating the contact angle hysteresis. Bormashenko et al.[20] installed a honeycomb-shaped oil-infused polycarbonate layer on electrodes as a dielectric layer to reduce the adhesive force between droplets and the dielectric layer. These setups can lower the applied voltage by eliminating the so-called pinning effect as the oil layer separates the droplet from contacting the substrate[29–31]. Lin et al.[21] used small droplets of 300 pL in an oil-filled microchannel (cross-section of 150 × 20 μm$^2$) and achieved a driving voltage of 7.2 V.

Among the methods listed above, droplet manipulation on oil-infused substrates or in oil-filled channels is more essential because the energy required to depin the droplet edge is much higher than that required to keep the droplet moving[32]. Moreover, eliminating the pinning effect is crucial to improve the reliability, especially when the driving voltage is limited because the driving force would not be sufficient for the depinning.

Droplet velocity is also a trade-off for the low-voltage actuation. He et al.[33] and Chang et al.[34] reported that the droplet velocity shows a power-function-like response to the applied voltage and it falls to $10^{-1}$ mm s$^{-1}$ when the applied voltage is less than 35 V on an oil-infused substrate[33] and 45 V in a parallel-plate hydrophobic channel[34]. However, from the perspective of the application, velocity of a few mm s$^{-1}$ is favorable for quick and reliable operation.

As stated above, satisfying both the low-voltage actuation and sufficient velocity ($\gtrsim 10^0$ mm s$^{-1}$) is essential to develop POC portable devices. Therefore, we investigate the motion of a droplet on an oil film manipulated by EWOD. We develop a physical model of the droplet velocity under lubrication. Furthermore, based on a lateral-view observation and a particle tracking velocimetry (PTV), we discuss the lubrication effect on the droplet velocity.

## II. METHODS

### A. Device overview

We develop four types of substrates shown in Fig. 1. These substrates have a common electrode pattern on a glass slide. As a non-lubricated EWOD substrate, we prepare Substrate A, which has a solid dielectric layer (CYTOP®), micropillars, and hydrophobic coating (Glaco Mirror Coat, Soft 99) on the electrodes and the micropillars. The lubricated EWOD substrate (Substrate B) is prepared by infusing silicone oil on Substrate A to form a lubricating film in between the micropillars. Note that the Glaco coating is performed to prevent the adhesion of the droplet to the micropillars. Note also that both the micropillars, Glaco coating, and silicone oil are dielectric materials and therefore act as dielectric layers. We further prepare Substrates C and D to observe effects of the solid dielectric layer. Substrate C removes the CYTOP® layer from Substrate B (*i.e.*, the dielectric layer consists of the micropillars, the Glaco coating, and the oil layer), whereas Substrate D removes the CYTOP® layer as well as the Glaco coating on the electrode from Substrate B (*i.e.*, the dielectric layer consists of the micropillars and the oil layer).

To ensure that the lubricating film stably exists beneath the droplet, we carefully chose sets of materials for the components to have positive Hamakar constants (*i.e.*, to induce van der Waals attraction force). When three components are interacting, the Hamakar constant can be derived by the Lifshitz theory with the static permittivity of each component[29,35]. Because the static permittivity is a function of the refractive index, we obtain conditions to have a positive Hamakar constant as combinations of the refractive indices. Namely, the Hamakar constant is positive when $n_e > n_l > n_d$[29], where $n$ denotes the refractive index and subscripts e, l, and d denote the electrode, the liquid layer, and the droplet, respectively. Under this condition, the film spontaneously stabilizes in nanometer scale owing to the van der Waals force. We chose gold or ITO (indium tin oxide) for the electrodes (ITO for the PTV measurement), silicone oil for the liquid layer, and pure water for the droplet to satisfy the condition. Furthermore, the lubricating film is stable as well in microscale, because the spreading factor S satisfies[28,29] $S > 0$, which was confirmed by measuring a contact angle of silicone oil against the solid with water as the third phase[28]: if the contact angle is zero, $S > 0$. We also confirmed that the droplets are fully covered by thin oil film[28,29,36]: we measured the diffusive evaporation time at room temperature with and without the silicone-oil layer on a substrate. On a substrate without oil layer, a water droplet

(10 μL) completely evaporates in ~5 minutes whereas a water droplet of the same volume survives more than 30 minutes on a substrate with oil layer.

We used five different oils of viscosity of 2–100 mm$^2$ s$^{-1}$ (Shin-Etsu Silicone, Table S1). We controlled the oil-layer thickness to pillar height by waiting for at least five minutes after pouring the oil on the substrate. After certain time, we confirmed that the layer settles down to the pillar height by gravity and the surface tension.

The contact angle and the contact angle hysteresis are 112 ± 4° and 59° on Substrate A and 76 ± 2° and <2° on Substrate B–D. Note that the contact angle hysteresis on Substrate A is relatively high for a hydrophobic-coated substrate because the coating thickness in this study is very thin (~50 nm).

DC voltage (≤ 75 V) was applied using a DC power supply (PLE-36-1.2, Matsusada Precision) and a power amplifier (HSA4101, NF Corporation).

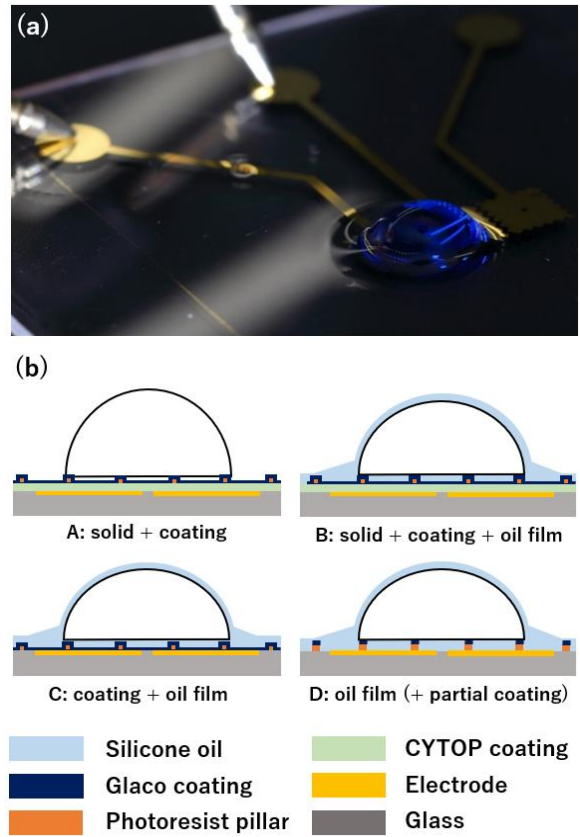

FIG. 1. (a) Photograph of the droplet migration on oil film by EWOD. (b) Schematics of Substrates A–D used in this study. Each substrate contains (A) solid dielectric, micropillars, and hydrophobic coating, (B) solid dielectric, micropillars, hydrophobic coating, and oil film in between the pillars, (C) micropillars, hydrophobic coating, and oil film in between the pillars, (D) micropillars and oil film in between the pillars (only top surface of the pillars is hydrophobic-coated). Glaco Mirror Coat was used for the hydrophobic coating.

## B. Device preparation

Devices were fabricated using the standard etching and lithography methods. We used different materials for the electrodes: gold electrodes for general operations and transparent indium tin oxide (ITO) electrodes for the particle tracking velocimetry (PTV) measurement. The gold electrodes were fabricated by the following procedure. First, a gold film was formed on a chrome layer (10 nm) by chemical vapor deposition (CVD). Second, the film was selectively etched using a photoresist (7790G-27cP, JSR Corporation) and an etchant (P3-19, Hayashi Pure Chemical for gold, and R8-16, Hayashi Pure Chemical for chrome). Then, the resist and residues were removed by cleaning the substrate with acetone, ethanol, and water. The fabrication procedure of the ITO electrodes was similar to that of the gold electrodes. An ITO-film-coated glass substrate (film thickness: 300 ± 20 nm, resistance: 5 ± 1 Ω / sq) was selectively etched by the photoresist and the etchant and cleaned with acetone, ethanol, and water. Figure 2a shows the design of the electrode panels. Each electrode panel has a size of 2.0 mm × 2.0 mm and comb-like perimeters. The gap between the electrodes was set to 0.1 mm. For Substrates A and B, a CYTOP® layer is spin-coated on the electrodes to form a solid dielectric layer of 1 μm thickness. The rotation rate for the spin-coating was 3 000 rpm, and the spin-coated substrates were heated first at 80◦C for 60 minutes (pre-baking) and then at 200◦C for 60 minutes (post-baking) to ensure the coating adhesion.

We fabricated micropillars using the standard photolithography method on the CYTOP® layer (Substrates A and B) or on the electrodes (Substrates C and D). The micropillars were made of the photoresist (7790G-27cP, JSR Corporation). We prepared a square-shaped pillar (14 μm × 14 μm) with three different inter-pillar spacings (14 μm, 21 μm, and 46 μm) and three different pillar heights h (0.38 μm, 0.75 μm, and 3.36 μm) (Fig. 2b–d). Note that three inter-pillar spacings correspond to the solid (pillar) fraction $\varphi$ = 25.0%, 16.0%, and 5.4%, respectively. The pillar size and the inter-pillar spacings were controlled by a photomask, and the pillar height was controlled by the rotation rate of a spincoater (500, 3 000, and 5 000 rpm for $h$ = 3.36 μm, 0.75 μm, and 0.38 μm, respectively). We performed a hydrophobic coating on the pillars with a Glaco Mirror Coat (Soft 99) solution to avoid the adhesion of the droplets on them[27,36–41]. Note that we performed the hydrophobic coating after the micropillar development (*i.e.*, coating on the pillar forest) for Substrates A–C. In contrast, we performed the coating before the development (*i.e.*, the resist layer is coated prior to developing the pillars) for Substrates D to selectively coat the top portion of the pillars. The coating thickness (~50 nm) was measured by a 3D optical profilometer (Profilm3D, Filmetrics).

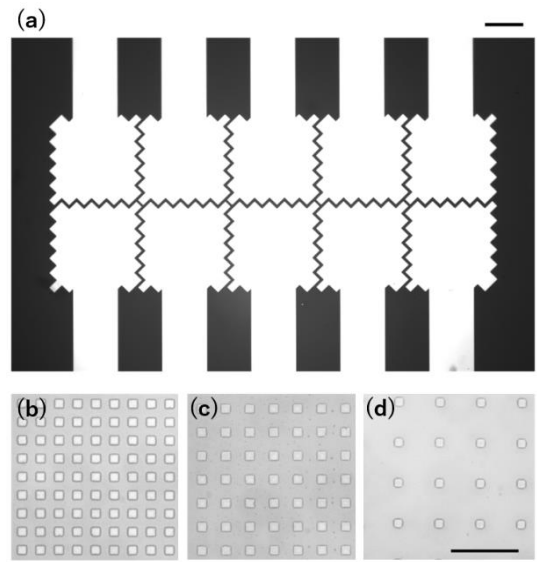

FIG. 2. (a) Design of the electrode panels. Panel size: 2.0 mm. Inter-panel gap: 0.1 mm. Scale bar corresponds to 1 mm. (b–d) Micrographs of the pillars. The pillars were fabricated by the standard photolithography method. Scale bar corresponds to 100 μm. All the pillars are square-shaped (14 μm × 14 μm) and their height is either $h$ = 3.36 μm, 0.75 μm, or 0.38 μm. Inter-pillar spacings are (b) 14 μm, (c) 21 μm, and (d) 46 μm, respectively.

## C. Velocity measurement

We capture the motion and the shape of the droplet by a high-speed camera (AX-100, Photron) at 2 000 fps from the side with a macro lens (25 mm F2.8 2.5-5X ULTRA MACRO, LAOWA) and a backlight illumination using an LED light source (KL 2500 LED, SCHOTT AG). The velocity of the droplet (as a body) is measured from the side-view images by measuring the center of the mass of the droplet.

We also perform the three-time-steps PTV to measure the velocity inside the droplet. Figure 3 shows a schematic of the experimental setup. Light from a mercury lump (INTENSILIGHT C-HGFI, Nikon) connected to an inverted microscope (Ti-2, Nikon) is filtered by an excitation filter (ET470/40x, Chroma Technology Corporation) and a dichroic mirror (T495lpxr, Chroma Technology Corporation) and illuminates fluorescent particles in a droplet through a 4× objective lens (CFI Plan Flour 4X, Nikon). The emission light from the fluorescent tracer particles (diameter: 10 μm, emission wavelength: 490 nm, excitation wavelength: 530 nm, G1000, Thermo Fisher Scientific) transmits the dichroic mirror and an emission filter (ET525/50m, Chroma Technology Corporation), and a high-speed camera captures the signal. The particle signal was captured at 2 000 fps and the shutter speed of 5 000 s$^{-1}$.

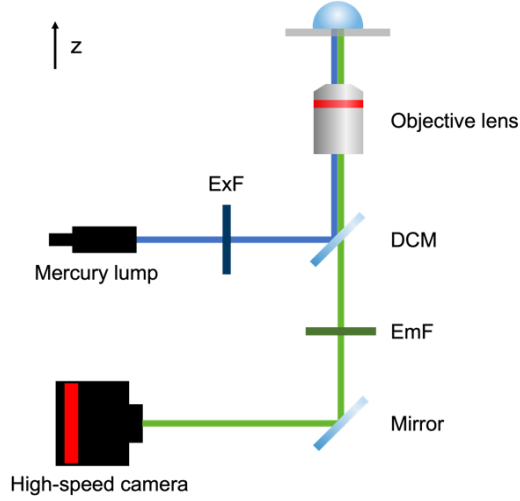

**FIG. 3. Schematic of the PTV setup. ExF: Excitation filter, DCM: Dichroic mirror, EmF: Emission filter.**

## III. PHYSICAL MODEL

In the following part we estimate the driving force and the resistance force of a moving droplet. In general, a driving force exerted on a static droplet is described by the surface tension and a difference in the receding and advancing contact angles $\theta_r$ and $\theta_a$ as[32]

$$F_{\text{drive}} = \gamma L k (\cos \theta_a - \cos \theta_r), \quad (1)$$

where $\gamma$, $L$, and $k$ denote interfacial tension, contact width of the droplet, and a numerical factor determined by the droplet shape, respectively. On the other hand, under the EWOD condition, the dielectrically induced change in the contact angle is expressed by Eq. (2)[1,12,42]

$$\cos \theta_E = \cos \theta_0 + \frac{\varepsilon_r \varepsilon_0}{2 d \gamma} E^2, \quad (2)$$

where $\theta_0$ and $\theta_E$ are contact angles before and after applying voltage, $\varepsilon_r$, $\varepsilon_0$, $d$, and $E$ denote relative permittivity of the dielectric layer, the permittivity of vacuum, thickness of the dielectric layer, and the applied voltage, respectively. With assumptions that $\theta_E \approx \theta_a$ and the change in θr is negligible at the initial stage of migration (i.e., $\theta_r \approx \theta_0$), we can derive a correlation of $F_{\text{drive}}$ (force to drive a droplet) and $E$ by combining Eqs. (1) and (2):

$$F_{\text{drive}} \approx \frac{R \varepsilon_r \varepsilon_0}{d} E^2, \quad (3)$$

where $R$ denotes droplet radius and $k = 1$, $L \approx 2R$ are assumed. Equation (3) is also valid during the motion when the substrate is oil-infused because the pinning effect is ruled out.

Subsequently, we estimate the drag force that should be balanced with $F_{\text{drive}}$. Under the lubricated condition, the drag to be considered are viscous frictions in (i) the thin liquid film, (ii) the foot of the droplet, and/or (iii) the oil meniscus[29,36,37]. In the following part we discuss these drag forces and derive equations that correlate the velocity and the applied voltage.

(i) In the thin oil film of thickness $h$, the viscous dissipation is proportional to oil viscosity $\eta_o$ and velocity profile $U_i / h$, where $U_i$ is the velocity at the oil–droplet interface. We assume that $U_i$ is constant at any point and is equals to the droplet velocity (as a body) $U$. Taking into account the effective area $\sim R^2$, the friction force in the film $F_{film}$ is described as

$$F_{film} \approx \frac{\eta_o U R^2}{h}. \quad (4)$$

(ii) The foot of the droplet deforms as the droplet migrates. As a result, a thin oil layer is formed according to the Landau–Levich–Derjaguin law[29,37]. The thickness of the layer $e$ is a function of the capillary number $Ca = \eta_o U / \gamma_{ow}$ ($\gamma_{ow}$ denotes oil–water interfacial tension) and can be estimated as $e \approx R Ca^{2/3}$. In case of the pillar array, the formation of the layer is expected only for $h < e$ (Seiwert et al.[43]), which leads a critical velocity[37] $U^* \approx (\gamma_{ow} / \eta_o)(h / R)^{3/2}$. If the droplet velocity exceeds $U^*$, the friction force at the foot of the droplet can be scaled as[29,37]

$$F_{foot} \approx 2\pi \gamma_{ow} R (\eta_o U / \gamma_{ow})^{2/3}. \quad (5)$$

(iii) In the oil meniscus, dynamic characteristics should be considered to estimate the viscous dissipation. The resulting friction force in the meniscus $F_{meniscus}$ is described as[36,37]

$$F_{meniscus} \approx \gamma_o^{1/3} \phi R (\beta \eta_o U)^{2/3}, \quad (6)$$

where $\varphi$ and $\beta$ denote the solid fraction and a numerical factor (detail of the model is given in Supplementary Information). We now correlate the above friction forces and the driving force to derive relationships between the droplet velocity and applied voltage. If the dissipation in the liquid film is dominant, the driving force $F_{drive}$ [Eq. (3)] should be equal to the friction force $F_{film}$ derived by Eq. (4). Therefore, the droplet velocity $U$ is given by

$$U \approx \frac{h \varepsilon_r \varepsilon_0}{d \eta_o R} E^2. \quad (7)$$

Keiser et al.
37
showed that the magnitudes of $F_{foot}$ and $F_{meniscus}$ are on the same order. Therefore, as we perform the scaling analysis, we only consider the case that dissipation in the meniscus is dominant. In this case, $U$ is given from Eqs. (3) and (6) as

$$U \approx \frac{1}{\beta \eta_o \gamma_o^{1/2}} \left( \frac{\varepsilon_r \varepsilon_0}{d \phi} \right)^{3/2} E^3. \quad (8)$$

These relationships suggest that the droplet velocity is proportional to the square or cubic of the applied voltage depending on the dominant region of the dissipation. Moreover, both Eqs. (7) and (8) predict that the velocity increases as $d$ decreases. Assuming that the permittivity of the solid dielectric layer (of thickness $d_{solid}$) and silicone oil are on the same order, we can consider $d = d_{solid} + h$, and maximum velocity is obtained at the limit of $d_{solid} \to 0$, *e.g.*, on Substrate D.

## IV. RESULTS AND DISCUSSION

### A. Validation of the physical model
In this section we compare droplet velocity on different substrates to validate the physical model proposed in the previous section. First, Substrates A and B (non-existence and existence of the lubricating film) are compared to validate Eqs. (3) and (4). Second, Substrates B–D (different thickness of the solid dielectric layer) are compared to confirm the effect of the solid layer thickness on the droplet velocity that are predicted by Eqs. (4) and (6).

We perform lateral-view observations to validate the model developed in III by comparing the droplet motion on Substrates A and B (viscosity of the infused silicone oil: 20 mm² s⁻¹). A 10-μL water droplet is manipulated at 75 V, a minimum voltage that can drive the droplet on Substrate A (Fig. 4). The maximum velocity on Substrates A and B are 29.3 mm s⁻¹ and 23.8 mm s⁻¹, respectively. The differences in the receding and advancing contact angles immediately before the droplets start moving are $\theta_r - \theta_a = 23.5°$ on Substrate A and 1.4° on Substrate B. The small contact angle difference in the oil-infused condition (resulting from the effective pinning elimination) is analogous to the low hysteresis of droplets on slippery liquid-like surfaces[36]. Note that the contact angle difference is almost constant on Substrate B for $E \lesssim 75$ V and the minimum voltage on Substrate B is 17 V.

From the observation results, we obtain two different sets of $F_{\text{drive}}$ [from the contact angle differences and Eq. (1)] and the minimum applied voltages that lead to the following relationship if Eq. (3) is valid (we assume the contact angle difference on Substrate B as constant, $d$ equals a sum of the pillar height and the CYTOP® thickness):

$$E_B \approx \left(\frac{F_B}{F_A}\right)^{1/2} E_A, \qquad (9)$$

where $E$ and $F$ denote the minimum applied voltage and the driving force, subscripts A and B denote Substrates A and B, respectively. Substituting the measured contact angles and $k = 1$ into Eq. (1), we obtain $F_A = 35$ μN and $F_B = 2.1$ μN. Substitution of these values into Eq. (8) predicts $E_B \approx 18$ V, which is excellently close to the measured value of 17 V. Therefore, we conclude that Eq. (3) is valid.

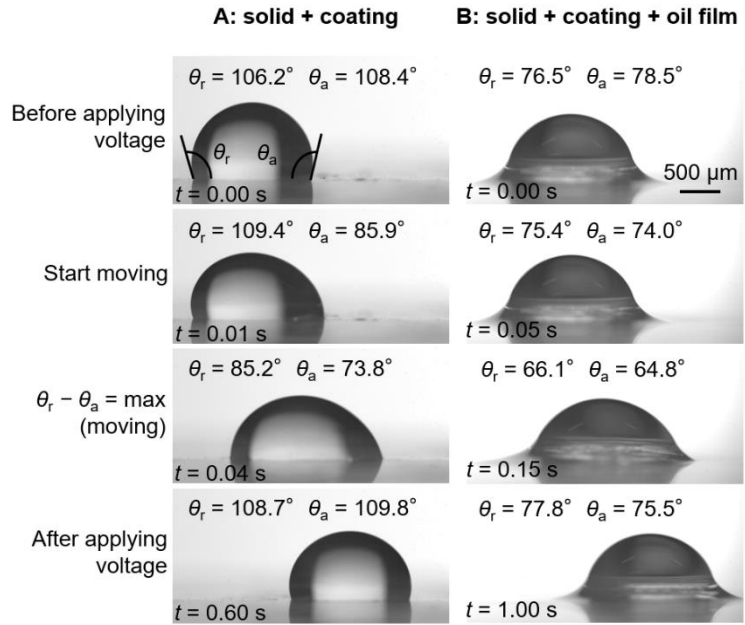

**FIG. 4. Side view of the droplet motion by EWOD on (left column) Substrate A and (right column) Substrate B with applied voltage of 75 V. Applied voltage, thickness of CYTOP® (1 μm), and micropillar height (0.75 μm) are in common, but a liquid film ($\eta_o = 20$ mm² s⁻¹) is infused for the right case. Scale bar corresponds to 500 μm.**

We further investigate the detail of the droplet migration. To understand the origin of the drag, we measure velocity profiles in three different horizontal planes inside the migrating droplets on Substrates A and B using PTV (Fig. 5 and Fig. S1). Figure 5a shows a typical velocity profile on

Substrate A obtained at $z = 100$ μm, where $z$ is an axis normal to the substrate and its origin locates at the droplet–oil interface. Note that the height of the droplets is approximately 0.8 mm on both Substrates A and B. It indicates that the velocity vectors are approximately uniform in the same plane except for those close to the perimeter, where three-dimensional flows exist, and error vectors due to the light reflection at the oil interface are significant on Substrate B. For that reason, we average velocity at the center region (30% of total area) and normalize them by the velocity of the center of mass of the droplet.

Figure 5a also indicates that the droplet shape is elongated in the migrating direction, whereas the shape is close to circle on Substrate B [The maximum change in the area of the droplet bottom plane (droplet–solid or droplet–oil interface) before and during the migration is +29% on Substrate A, whereas it is +8% on Substrate B, see also Fig. S1]. The deformation of the perimeter suggests that high shear force is exerted on the foot of the droplet. Figure 5b shows averaged velocity profiles in the $z$ direction normalized by the droplet migration velocity ($U^* = U_{plane} / U$, where $U^*$ and $U_{plane}$ denote the normalized velocity and measured plane velocity, respectively). The velocity profiles show that all the velocities in different $z$ planes are almost the same as the droplet velocity on Substrate B, whereas the velocity near the electrode (substrate) is faster than the body on Substrate A. This result implies that the oil infusion effectively reduces the drag caused by high shear due to the large velocity gradient at both substrate–droplet boundary and inside the droplet. Consequently, droplets can migrate as a body and save energy loss by deformation and drag. The results also imply that the droplet–oil interface is effectively driven on Substrate B by generating lubricating flow inside the oil film, and therefore the film not only eliminates the pinning effect but also assists the motion of the droplet. These results suggest that the velocity of the lubrication flow is significantly higher than the case of droplets sliding on inclined LISs where the interface velocity is estimated as[37] $U_i \approx U\eta_w h / (\eta_o R)$.

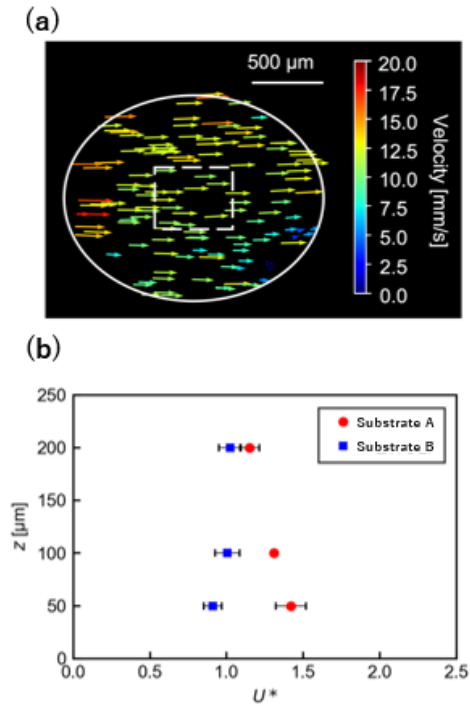

**FIG. 5. (a) Velocity vector distributions in a droplet migrating from left to right on Substrate A ($z = 100$ μm, $E = 75$ V. $z = 0$ indicates the droplet bottom). The solid white line indicates the perimeter of the droplet. The dashed white rectangle indicates the velocity-averaging area (30% of total area). Each vector is subtracted by the droplet velocity. (b) Velocity profiles in the z direction on Substrates A (red circles) and B (blue squares) ($E = 75$ V). The velocity is shown as the velocity relative to the droplet motion.**

Whatever the dominant drag force is, both Eqs. (7) and (8) indicate that the droplet velocity at a fixed voltage depends on $d$. Furthermore, Eq. (7) predicts the ratio $h / d$ is also important if $F_{film}$ is dominant. To check the effects of $d$ and $h/d$, we compare the droplet velocity at 75 V on Substrates B ($d \approx 1.75$ μm, $h/d \approx 0.42$), C ($d \approx 0.80$ μm, $h/d \approx 0.94$), and D ($d \approx 0.75$ μm, $h/d \approx 1$). Consequently, the average velocities on the respective substrate are 21.9 mm s$^{-1}$ (Substrate B), 27.2 mm s$^{-1}$ (Substrate C), and 27.1 mm s$^{-1}$ (Substrate D) (droplet velocity on Substrate A at 75 V is 29.3 mm s$^{-1}$). The result suggests that removing of the solid layer effectively increases the droplet velocity, but the effect is less than predicted: it would be because the film thickness became thicker than expected by the droplet motion[29,37]. McHale et al.[27] reported that the height of the oil wetting ridge around a droplet shows dependency on the oil-film thickness. With 20 cSt silicone oil, the ridge height around a small water droplet is ~400 μm when the oil thickness is 21.1 μm, whereas it is less than 50 μm on a 3.1-μm-thick film. It and Fig. 4 imply that the oil thickness in this study could have ~10 μm. However, it is not critical if $F_{film}$ is the dominant friction force because Eq. (7) predicts the film thickness does not affect the motion if the lubrication approximation is valid and $h/d \approx 1$.

## B. Characteristics of the droplet velocity

To understand the droplet migration mechanism, we measure the velocity of a 10-μL droplet manipulated on Substrate C with various applied voltage $E$ and oil viscosity $\eta_o$ under the condition of $\varphi$ = 25.0% and $h$ = 0.75 μm. Figure 6a shows the relationship between the droplet velocity $U$ (maximum velocity of the center of the mass) and $E$. The result shows that $U$ is proportional to $E^2$, indicating that the dissipation in the oil film is dominant and Eq. (7) explains the motion. It indicates that the mechanism of the droplet migration on lubricated substrates is different from the mechanism of a sliding droplet on lubricated substrates where the dominant resistance force is either $F_{foot}$ or $F_{meniscus}$ and $U \approx E^3$ relationship dominates the motion for the velocity range of this study[29,37].

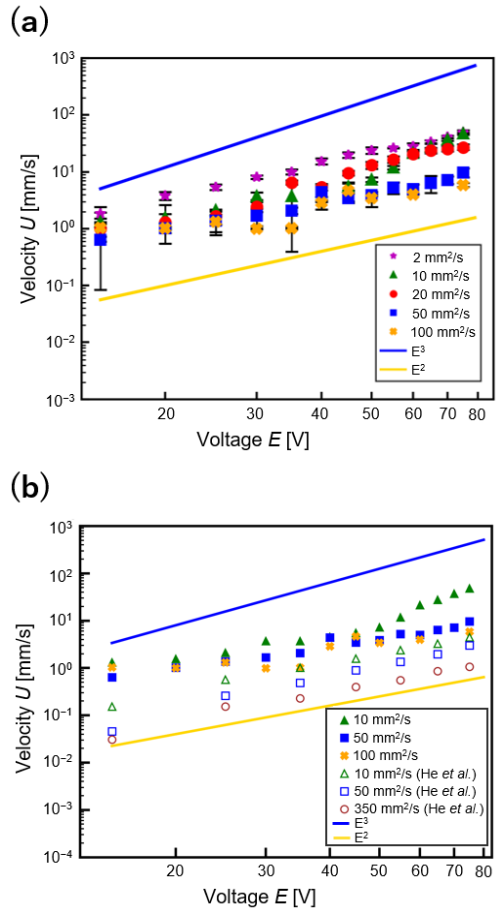

FIG. 6. (a) Relationship between the droplet velocity $U$ and the applied voltage $E$ with different oil viscosity $\eta_o$ on Substrate C ($h$ = 0.75 μm, $\varphi$ = 16.0%). (b) Comparison of the $U$–$E$ relationship in this study (on Substrate C, $h$ = 0.75 μm, $\varphi$ = 16.0%: filled symbols) and that in He et al.[33] (on porous-structured LIS: open symbols). Blue and Yellow solid lines indicate trends of $E^3$ and $E^2$, respectively.

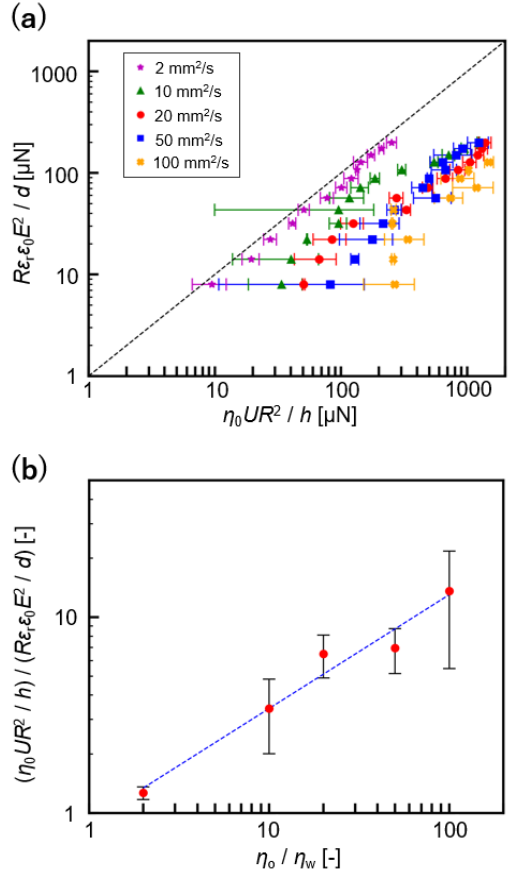

FIG. 7. Effects of the oil viscosity on Substrate C ($h$ = 0.75 μm, $\phi$ = 16.0%) in the relationship between $F_{\text{drive}}$ and $F_{\text{film}}$. (a) Comparison of $R\varepsilon_r\varepsilon_0 E^2 / d$ ($\approx F_{\text{drive}}$) with $\eta_o U R^2 / h$ ($\approx F_{\text{film}}$). (b) Balance of the resistance force and the driving force ($\eta_o U R^2 / h$) / ($R\varepsilon_r\varepsilon_0 E^2 / d$) ($\approx F_{\text{film}} / F_{\text{drive}}$ for various oil / water viscosity ratio $\eta_o / \eta_w$. Blue dashed line indicates a fitting curve ($\eta_o U R^2 / h$) / ($R\varepsilon_r\varepsilon_0 E^2 / d$) = $0.8969(\eta_o / \eta_w)^{0.5809}$.

As discussed in the previous section, the relative thickness of the lubrication layer to the solid (unmovable) dielectric layer affects the droplet velocity. To confirm this effect, we compare our results and the results on a LIS-type surface by He et al.[33]. Figure 7 shows the droplet-velocity dependence on the applied voltage on Substrate C (filled symbols, $\eta_o$ = 10, 50, and 100 mm² s⁻¹) and on the LIS=-type surface by He et al.[33] (open symbols, $\eta_o$ = 10, 50, and 350 mm² s⁻¹). It indicates that the droplets on Substrate C migrate faster than those on the surface by He et al.[33], while the $U \approx E^2$ trend is held on both substrates. The difference is again explained by Eq. (7): the effective thickness of the lubrication layer (a thickness of a region where the lubricant can freely flow to form a lubrication flow) affects the drag. On the surface by He et al.[33], a large part of the lubricant is immobilized in the porous structure and the effective thickness of the lubrication layer decreases significantly. We estimate a typical value of $h / d$ for porous-structured substrates to be $10^{-1}$ and it decreases the velocity by a factor of $10^{-1}$ in comparison with the velocity on Substrate C (pillar-structured surface, $h / d \approx 1$).

Figure 7a shows a comparison of the magnitudes of $R\varepsilon_r\varepsilon_0 E^2 / d$ ($\approx F_{\text{drive}}$) and $\eta_o U R^2 / h$ ($\approx F_{\text{film}}$) for various $\eta_o$. The diagram indicates that the driving force and the drag force are in a linear relationship and that $F_{\text{film}}$ is the dominant drag force. Furthermore, we confirmed that $F_{\text{film}} / F_{\text{meniscus}} > 10$ for all the cases shown in the diagram. However, Fig. 7a also indicates that $F_{\text{drive}}$ is smaller than $F_{\text{film}}$ (the black dashed line indicates 1 : 1 line), particularly for highly-viscous oils. Figure 7b shows a relationship between the oil viscosity (normalized by the water viscosity $\eta_w$) and a ratio of two forces ($\eta_o U R^2 / h$) / ($R\varepsilon_r\varepsilon_0 E^2 / d$) ($\approx F_{\text{film}} / F_{\text{drive}}$). The relationship shows an exponential correlation $F_{\text{film}} / F_{\text{drive}} \approx (\eta_o / \eta_w)^{0.58}$. This relationship implies that the actual viscous resistance $F_{\text{film,act}}$ is smaller than expected ($\eta_o U R^2 / h$) and the gap between them exponentially increases with $\eta_o / \eta_w$. It could be due to the droplet internal flow (rolling motion of droplets), which is omitted in our physical model: in the model we assume the velocity profile (in $z$ direction) inside the droplet is uniform, but it could have a gradient as in the case of a droplet sliding down on lubricated substrates[29,36,37].

The rolling motion of a droplet can reduce the velocity at the interface of the droplet bottom and the lubrication layer and $F_{\text{film,act}}$ is reduced as a consequence. Although the mechanism to generate the rolling motion is yet to be investigated, this hypothesis is plausible from a perspective of the viscous resistance inside the droplet. The viscous resistance by the shear inside the droplet can be estimated by Eq. (10):

$$F_{\text{drop}} \approx \eta_w \frac{dU}{dz} R^2, \quad (10)$$

where $dU/dz$ is the velocity gradient in the droplet. Because the maximum $F_{\text{drop}}$ can be estimated by taking $dU/dz$ from the velocity profile on Substrate A (because lubrication suppresses the inner flow, see Fig. 5), the effect of the inner flow can be evaluated by comparing $F_{\text{drop}}$ on Substrate A and $F_{\text{film}}$ for low oil viscosity. As a result, $F_{\text{drop}}$ on Substrate A (at 75 V) is calculated as $10^{-2}$ µN and $F_{\text{film}}$ for $\eta_o = 2$ mm$^2$ s$^{-1}$ is calculated as $10^{-1}$ µN with $U = 0.1$ mm s$^{-1}$, which is two order smaller than the actual migration velocity. This result implies that the energy loss inside the droplet is negligible while the rolling motion affects the velocity in the lubrication layer resulting in the reduction of $F_{\text{film,act}}$.

We discussed that the dominant friction force is $F_{\text{film}}$ for all applied voltage ranges in this study and therefore the droplet velocity is scaled by $E^2$. In addition to the $E^2$ correlation, we can expect that $U$ increases with $R^{-1}$ as $U$ obeys Eq. (7). Figure 8a shows a dependence of $U$ on $R$ for $\eta_o = 20$ mm$^2$ s$^{-1}$ and $E = 75$ V. The diagram shows an increasing trend of $U$ with decreasing $R$. Moreover, Fig. 8b shows that $UR$ for different $R$ take almost a constant value. It implies that smaller droplets have higher mobility as we expected.

In contrast to Eq. (8) and instinct, Eq. (7) predicts that $U$ does not depend on $h$ when $h \approx d$ is satisfied. Figure 9a shows $U$ as a function of $E$ for different $h$. The result shows a slight increase in $U$ with $h$, but the difference is not significant, at least in this range of $h$. On the other hand, the effect of $\varphi$ is non-negligible (Fig. 9b). In these three cases, $U$ takes the maximum with the middle of the fraction $\varphi = 16.0\%$. This could be explained as follows: the frictional loss and the viscous dissipation becomes smaller as $\varphi$ decreases, but it turns back at a particular point, under which drag increases as the bottom of the droplet deforms, and the pillars act as bumps[45].

## C. Minimum activation voltage under lubrication

Finally, we perform the droplet manipulation with applied voltage as low as possible. As mentioned earlier, the minimum voltage that can drive the droplet with a droplet volume $V = 10$ µL and $\eta_o = 20$ mm$^2$ s$^{-1}$ on Substrate B is 17 V, and it decreased to 13 V on Substrates C and D. He et al.[33] reported the minimum voltage on their LIS-type surface with similar condition was 8 V with $\eta_o = 10$ mm$^2$ s$^{-1}$, $V = 10$ µL, which implies the liquid-layer thickness does not affect the start of the motion. However, the minimum voltage is decreased by decreasing the droplet radius and the oil viscosity. We confirm that a 10-µL droplet migrates with $E = 6$ V (Fig. 10) and a 5-µL droplet migrates with $E = 5$ V with $\eta_o = 2$ mm$^2$ s$^{-1}$, $\varphi = 16.0\%$, and $h = 0.75$ µm. In both cases the droplet velocity was ~0.1 mm s$^{-1}$.

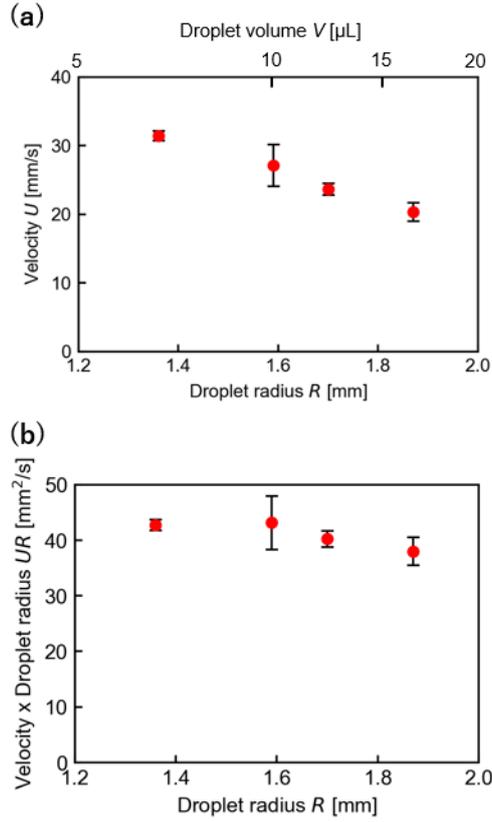

**FIG. 8.** Droplet velocity for different droplet radius R on Substrate C ($h = 0.75$ μm, $\varphi = 16.0\%$, $\eta_o = 20$ mm$^2$ s$^{-1}$, $E = 75$ V). (a) $U$ vs. $R$. (b) $UR$ vs. $R$.

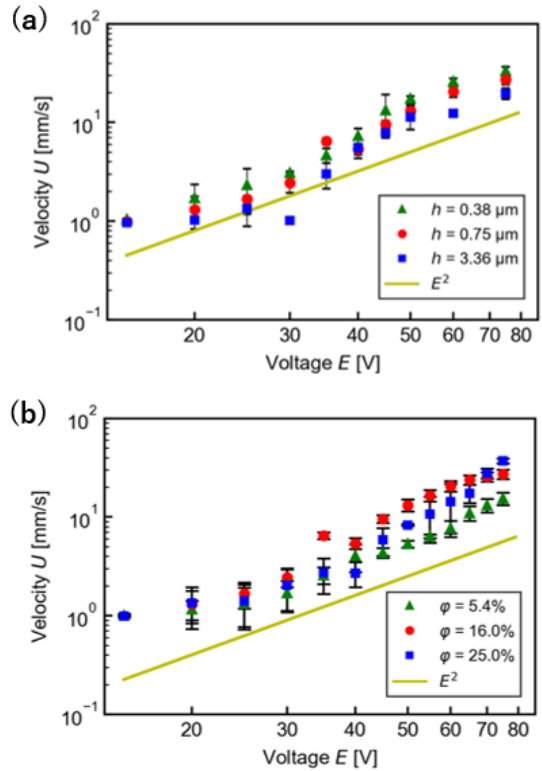

**FIG. 9.** Dependence of the droplet velocity on substrate geometries on Substrate C ($\eta_o = 20$ mm$^2$ s$^{-1}$, $E = 75$ V). (a) Effects of the thickness of the dielectric liquid $h$ ($\varphi = 16.0\%$). (b) Effects of the solid fraction $\varphi$ ($h = 0.75$ μm). Yellow solid line indicates a slope of $E^2$.

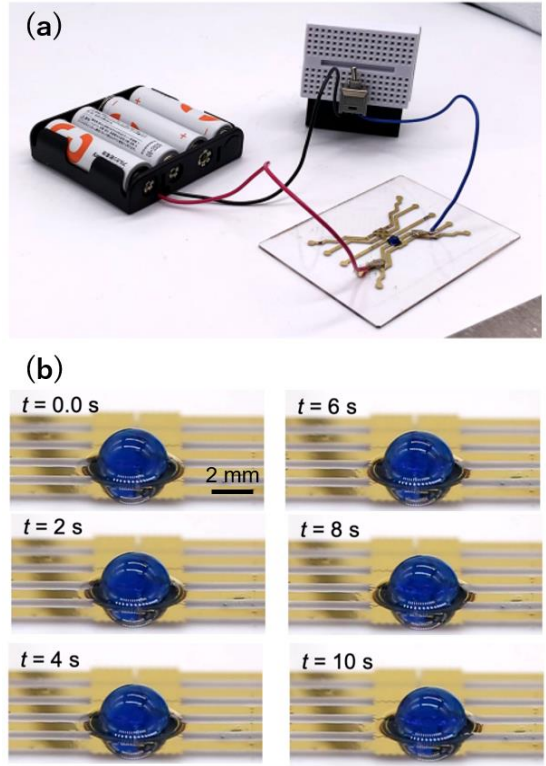

FIG. 10. Low voltage actuation. (a) Experimental setup. The electrodes on Substrate C are connected to four AA batteries and an on/off switch. (b) Successive images of the droplet motion ($h$ = 0.75 μm, $\varphi$ = 16.0%, $\eta_o$ = 2 mm$^2$ s$^{-1}$, $V$ = 10 μL, $E$ = 6 V).

## V. CONCLUSIONS

We developed micropillar-arrayed EWOD substrates to investigate the droplet motion on a lubricating film manipulated by EWOD. The developed substrates can retain silicone oil in between the micropillars by capillarity and can be used as lubricated EWOD substrates. The infused oil effectively eliminates the pinning effect so that droplets migrate as low as 5 V. Furthermore, because of the sparse micropillars, lubrication flows of oil are generated once a driving force is exerted on the droplet contacting with the lubricating film.

  The lubrication effect brings high velocity ($\geqq$ 1 mm s$^{-1}$) at the applied voltage of 15 V, approximately tenfold the previous study on a liquid-infused surface containing porous structures[33]. It is achieved by mobilizing the lubricating film and reducing the immobilized layer. A physical model developed in this study can predict these behaviors well. The model also predicts that the lubricating-film thickness does not affect the motion unless the lubrication approximation is valid and the oil-layer thickness is the same or close to the total dielectric-layer thickness ($h / d \approx 1$).

  We developed a physical model of the droplet motion based on a scaling analysis. Several different drag forces were considered. As a result, different relationships between the velocity $U$ and the applied voltage $E$ depending on the dominant drag force were obtained. We then obtained $U \approx E^2$ relationship from experimental observations. This relationship indicates that a drag force induced by the lubrication flow is a dominant resistance. However, a balance between the driving force and the drag force predicted by the model was affected by the oil viscosity. It is likely due to the internal motion of droplets.

  The physical model also predicts that the droplet velocity increases when small droplet is manipulated. It is because the drag force exerted on the droplet is a function of the droplet radius. We confirmed that reducing of the droplet radius (as well as the oil viscosity) is also effective in lowering the minimum driving force. Finally, manipulation of a 5-μL droplet by 5 V was performed: it opens a door for manipulations in various ways; for instance, coupling with photoelectrowetting[38–41] may lead to droplet handling on a smartphone, novel adjustable lenses, and portable lab-on-a-chip devices.

## VI. SUPPLEMENTARY MATERIAL

Additional information of the oil properties, the physical model, and the velocity measurements are given in Supplementary Material.

## ACKNOWLEDGMENTS

The work was supported by the JSPS KAKENHI (Grant No. 20K14672). KY appreciates Kazuma Yoda (Tokyo University of Science) and Keiko Ami (Tokyo University of Science) for their technical support. KY also appreciates Hiroaki Katsuragi (Osaka University) for helpful discussions.

## DATA AVAILABILITY STATEMENT

The data that support the findings of this study are available from the corresponding author upon reasonable request.